\documentclass[twocolumn,aps,floats,floatfix]{revtex4}
\usepackage{graphicx,amssymb}

\newcommand{\beq}{\begin{equation}}
\newcommand{\eeq}{\end{equation}}

\newcommand{\bea}{\begin{eqnarray}}
\newcommand{\eea}{\end{eqnarray}}

\newcommand{\ba}{\begin{array}}
\newcommand{\ea}{\end{array}}

\newcommand{\nuke}[1]{}
\begin{document}
\title{ Coexistence of Bose condensation and pairing in Boson mixtures }
\author{S. T. Chui}
\affiliation{Bartol Research Institute and Dept. of Physics and Astronomy,\\
University of Delaware, Newark, DE 19716}
\begin{abstract}
We consider the problem when there are two kinds of Bosons with an attraction between them.  We find the system to consist of two Bose condensates with an additional pairing order between the Bosons.  The properties of this state are discussed.
\end{abstract}
\maketitle

\section{Introduction}
In general if there is an attraction between two particles they can form a bound state if the kinetc energy cost is lower than the potential energy gained. This is true for both Fermions and Bosons.  For a many particle system, the two particle bound state becomes a collective state. Because of the identity of particles, bound states can be formed between any two particles that are attracted to each other. The signature of this collecive state is the pairing order parameter. For Fermions, near the Fermi energy the kinetic energy cost is always low enough, as is manifested in a log divergent two particle response function. For Bosons, the kinetic energy cost is higher, but because there is no exclusion principle, each particle can interact with many other particles and the interaction is enhanced. We thus expect a collective "paired" state to also form if the interaction is attractive.
 
There has been much studies of the Bose-Einstein condensation and of superconductivity of Fermions but there were no previous system in which Bose condensation and Boson pairing occur at the same time. This possibility has been considered for a one component attractive Bosons\cite{No} but unfortunately one component attractive Bosons are unstable against collapse.\cite{Hulet} 
In this paper we consider a mixture of two Bosons such that the intra-component interaction is repulsive but the intercomponent interaction is attractive. We find that, when the intracomponent repulsion is strong enough, the system  consists of two Bose condensates with  pairing  between the Bose atoms of the two different components.   This state provides for an opportunity to study the new physics of the coupling of the superfluid order and the pairing order. An experimental realization of this system can be found as a mixture of cold $^{87}Rb$ and $^{41}K$ atoms. The interaction between Rb and K can be controlled by a Feshbach resonance.  The physical properties of this state and different ways to experimentally detect it are discussed. 

 
There are usually two limits to describe the collective "bound" state, depending on whether the interparticle spacing is larger or smaller than the size of the bound pair. In the latter case, a mean field approximation is appropriate. Again this idea applies to both Fermions and Bosons. The mean field description of the pairing for Fermions has been discussed by BCS and the basic mathematical manipulation can be carried over. A major difference between Fermions and Bosons is the appearance of the condensates for Bosons and it is necessary to treat the zero momentum states separately. It is also necessary include the interaction of the condensate and the pair function. 

\section{Formulation}
We first derive the basic excitation and "gap" equations for this state. 
We start with the Hamiltonian
of the particles of mass $m_i$ with wave function
$\Psi_i$ under a trapping potential $V_i$:
The Hamiltonian $H_t$ is given by
\beq
H_t=\int d^3r\sum [V_i|\Psi_i|^2
+\hbar^2|\nabla \Psi_i|^2/2m_i 
\eeq
$$
+
0.5 \sum_{i, j} G_{ij}\Psi_{i}^+\Psi_j^+\Psi_{j}\Psi_i]$$
As usual $G_{ij}=2\pi\hbar^2 a_{ij}/\mu_{ij}$ where $a_{ij}$ is the scattering length and $\mu_{ij}$ is the reduced mass.
The Fourier  transform of the wave function can be written as
$\Psi_i^+(r) =\sum_k \exp(i{\bf k\cdot r})a_{i,k}^+/\sqrt{V}$ where
$a_{i,k}^+$ is the particle creation operator of momentum k for component i and $V$ is the volume. 
Following BCS 
we define an order parameter $D=\sum_{k\neq 0} < a_{1k}^+a_{2,-k}^+>/V,$ 
which can be determined together with the averages $D_i=\sum_{k\neq 0} < a_{ik}^+a_{i,-k}^+>/V$ self consistently via "gap" equations described below. The correlation function $D_i$ have been previously discussed
for the one component case\cite{Mahan}.

We first consider the case where the trapping potential is absent. We assume that the Bosons form condensates with condensate density $N_i$.
Following BCS and Bogoliubov, in mean field $H_t$ is approximated by a Hamiltonian given by a sum of the Hamitonian of each of the components and an interaction term between them:
\beq
H=\sum_{k\neq 0,i}\omega_{i0}a_{ik}^+a_{ik}+ 0.5(v_i+s_i)(a_{ik}a_{i,-k}+a^+_{i,-k}a^+_{ik})
+H_{int} .
\label{bog}
\eeq
where we sum over the component i and the wave vector k except for k=0. $\omega_{i0}=e_{ik}+v_i,$ $e_{ik}=\mu+\hbar^2 k^2/2m_i$ is the kinetic energy of component i, $v_i=N_iG_{ii},$ $s_i=G_{ii}D_i$ is a measure of how much atoms of the same species are not on top of each other;
\beq
H_{int}=0.5\sum_{k\neq 0} [u(a_{1k}a_{2,-k}+a^+_{1,-k}a^+_{2k})
\label{hint}
\eeq
$$
+w (a_{1k}^+a_{2,k}+a_{1,k}a^+_{2k})],
$$
$w=2(N_1N_2)^{1/2}G_{12}$ represent interaction with the Bose condensates, $u=u'+w,$ 
$u'=2G_{12}D$  represent the interaction with the pairing order. 
$H_{int}$ is motivated by the BCS theory.
We have assumed that the phases of $D$, $D_i$ are fixed and set them to be real numbers. Just as in the BCS theory, the hydrodynamics modes described by the Josephson equations\cite{phase} involve the change of phases of D and and $D_i$ and thus are not included in the present calculation.  Usually, $D_i<0$ and $D>0$. 

As usual, we perform the Bogoliubov transformation and define new eigen-operators 
$$\alpha_{j,k}=\sum_i c_{i,j}a_{i,k}-d_{ij}a^+_{i,-k},$$
with commutators given by
$$
[\alpha_i,\alpha_j^+]=\sum_l (c_{l,i}c^*_{l,j}-d_{l,i}d^*_{l,j})=\delta_{i,j}.
$$

From the conditions $[\alpha^+,H]=-\lambda \alpha,$
$[\alpha,H]=\lambda \alpha^+,$ we obtain the eigenvalue equation
$$
\hat H\left(
\begin{array}{c}
 c\\
 d\\
\end{array}
\right)=
\left(
\begin{array}{cc}
 \lambda & 0\\
 0&-\lambda\\
\end{array}
\right)
\left (\begin{array}{c}
 c\\
 d\\
\end{array}
\right)
$$
where
$$
\hat H=\left(
\begin{array}{cc}
 H_1 & H_2\\
 H_2&H_1\\
\end{array}
\right),\ \  
$$

$$
H_{1}=\left(
\begin{array}{cc}
 {e_1}+{v_1} & w\\
 w&{e_2}+{v_2}\\
\end{array}
\right),
H_2=\left(
\begin{array}{cc}
{v_1+s_1} & u \\
 u & {s_2}+{v_2}  \\
\end{array}
\right) 
$$
The setails of the solution of this equation is described in the Appendix.

\begin{figure}
\includegraphics*[width=6cm,height=6cm]{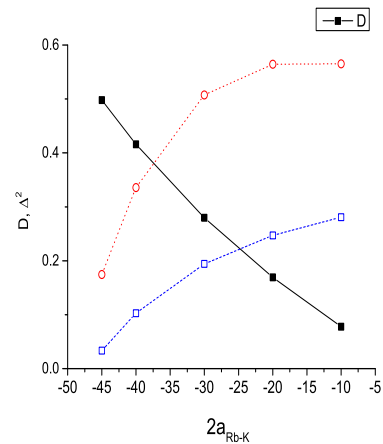}
\caption{The order parameter $D$ (black) in units of the trap volume density and the excitation energies 
$\Delta^2$ (blue $\square$ and red $\circ$ ) in units of the trapping energy as a function of the K-Rb scattering length in units of the Bohr radius} \label{ssfx}
\end{figure}
 
The gap equation becomes
$$
D=\sum_{i,k\neq 0}(c_{1,i}d_{2,i}+c_{2,i}d_{1,i})/(2V)
$$
The equations for the other parameters are
$$
D_j=\sum_{i,k\neq 0} c_{j,i}d_{j,i}/V
$$
The gap equations can be solved numerically. We discuss this next.

\section{Numerical illustration}
As an example we assume that the condensate density of the two components are the same and is equal to the central density of $10^6$ K atoms trapped in a pancake shaped trap of aspect ratio $\sqrt{8}$ and a trapping frequency of 100Hz. The scattering lengths are $a_{K-K}=66a_B$, $a_{Rb-Rb}=94.3 a_B$ where $a_B$ is the Bohr radius.
The result for $D$ in units of the trap volume density and $\Delta^2$ in units of the $(trapping\ \ energy)^2$ are shown in fig. (\ref{ssfx}) as a function of the K-Rb scattering length $a_{K-Rb}$ in units of the Bohr radius. 
That solutions to the gap equation exist shows that indeed the kinetic energy cost is low enough and the coexistence of the two order is indeed possible. As the Rb-K scattering length becomes more negative $D$ increases.

\begin{figure}
\includegraphics*[width=6cm,height=6cm]{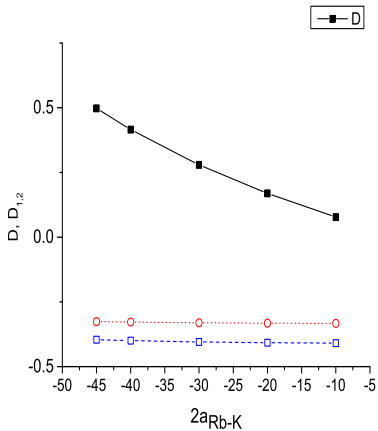}
\caption{The order parameter $D$ (black) and the parameters $D_K$ (blue $\square$ ) and $D_{Rb}$ (red $\circ$) in units of the trap volume density as a function of the K-Rb scattering length in units of the Bohr radius.} \label{ssfd}
\end{figure}

The parameters $D_i$ are not strong functions of $a_{Rb-K}$.  These, together with the order parameter, are shown in fig.(\ref{ssfd}). As is well known\cite{FW}, their finite values indicate  depletions of the condensate densities, $\delta N_i=\sum_{j,k\neq 0} d_{i,j}^2/V$, which is larger for more negative $a_{Rb-K}$. This depletion is small. For example, the changes $\delta N_{Rb}$ and $\delta N_{K}$ are equal to 2.35\% and 2.45\% respectively for $2a_{Rb-K}=-45 a_B$ .

For Fermions, there is a BCS to BEC transition as the attraction becomes strong. It is of interest to study the corresponding situation for the Boson case. In the study of one component attractive Bosons, the cloud collapses when the number of Bosons is large enough\cite{Hulet}
There is a resemblance between that problem and the present one.  Thus it is meaningful to ask if or when this state will also collapse? One can speculate that it may depend on the relative magnitude of $G_{11}$, $G_{22}$ vs $G_{12}.$ 
The energy $\Delta_-$ in eq. (\ref{eqgap}) becomes imaginary and the state is unstable
if $\alpha>0$. We found that this happens, for example, when there is no intra-component repulsion and
$G_{ii}=0.$ 
Our numerical result in fig. (\ref{ssfx}) shows that $\Delta^2$  decreases towards zero as $a_{K-Rb}$ is decreased. This is reasonable as we expect when the intra-component repulsion is not sufficient to prevent the collapse, an instability will occur. 
\begin{figure}
\includegraphics*[width=6cm]{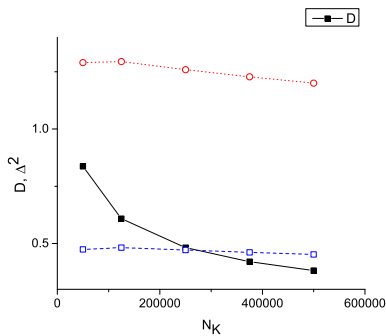}
\caption{The order parameter $D$ (black) in units of the trap volume density and the excitation energies 
$\Delta^2$ (blue $\square$ and red $\circ$) in units of the trapping energy as a function of $N_K$ with $N_{Rb}=2N_K$.} \label{ssfN}
\end{figure}

We have also investigated the dependence of our results on the superfluid density. An example of this is illustrated in fig.(\ref{ssfN}).
We assume that the Rb condensate density is twice that of the K condensate density which  is equal to the central density of $N_K$  atoms trapped in a pancake shaped trap of aspect ratio $\sqrt{8}$ and a trapping frequency of 100Hz. As the Boson density is decreased, the effective repulsion is decreased and $\Delta^2$ is increased. This shows that the existence of the pairing order does not depend on the existence of the superfluid. Physically, the pairing Hamiltonian $H_{int}$  remains finite even when $N_1=N_2=0$ because there is a  term proportional to $G_{12}D$.


\section{Pairing and Bose condensatation}
The additional pairing order makes the superfluid phase more favorable. We illustrate this with an example next.
There has been recent interest in the insulator-superfluid transition of Bosons in the presence of a periodic potential.  We found that in mean field, for the two component system in the presence of an external periodic potential, the pairing makes it more difficult to achieve the insulator phase. As usual we consider the Bose-Hubbard Hamiltonian $H=\sum_i H_{iS}+H_{iA};$
\begin{equation}
\label{q22}H_{iS}=\sum_{\alpha }\hat \lambda _{s\alpha} \hat N_{\alpha i}(\hat N_{\alpha i}-1)-\mu_{\alpha}\hat N_{\alpha i} 
\end{equation}
\begin{equation}
\label{q8}H_{iA}=\sum_{\alpha <ij>}-J_{\alpha} \hat a_{\alpha i}^{+}\hat a_{\alpha j}+\lambda _a \hat N_{1 i}\hat N_{2 i}
\end{equation}
where $\hat N_{\alpha i}= \hat a_{\alpha i}^{+}\hat a_{\alpha i}$ is the
number of atoms of component $\alpha$ at lattice site $i$ , $\lambda$ corresponds to the interaction strength and $J_{\alpha}$
is the hopping matrix element between adjacent sites $i$, $j$. We investigate the condition such that  component 1 is in the superfluid state while  component 2 is close to the insulating phase.
Let us introduce the superfluid order parameter\cite{Oost01} 
$\psi _2 =<\hat a_{2 i}^{+}>=<\hat a_{2 i}>$ and construct a
consistent mean-field theory by substituting 
\begin{equation}
\label{q111} \hat a_{2 i}^{+}\hat a_{2 j}
= \psi_2 (\hat a_{2 i}^{+}+\hat a_{2 j})-\psi _2 ^2 . 
\end{equation}
Then Eq. (\ref{q8}) yields the following $\psi_2$ dependent terms:
\begin{equation}
\label{q112}H_{iA}'=zJ [\psi _2 ^2- \psi _2 (\hat a%
_{2 i}^{+}+\hat a_{2 i})]+\lambda _a \sqrt{N_{1}}\psi_2 D , 
\end{equation}
where $z=2d$ is the number of nearest-neighbor sites, $d$ is the space
dimension. The last term is the mean field approximation of the interparticle interaction term. Central to this approximation is that, as is pointed above and illustrated in fig. (\ref{ssfN}),  the pairing order parameter $D$ remains finite even as the superfluid density $N_2$ becomes small. 
The last term in this Hamiltonian is absent in previous calculations of the one component case. It stabilizes 
the superfluid phase and makes it more difficult to become a Mott insulator. To summarise, there is a coupling term of the form $H_c=G_{12}(N_1N_2)^{0.5}D$ between the superfluid densities $N_{1,2}$ and the pairing order parameter $D$. Because the existence of the pairing order $D$ does not depend on the existence of the superfluid, $D$ reinforces $N_{1,2}$.

In the presence of a trap, $\Psi(r)=\psi_0(r)+\psi(r)$ where $\psi_0(r)$ corresponds to the Bose condensate
function. For the current experimental systems, the Thomas-Fermi approximation works well and we assume that 
is the case here. A key point is that the magnitude of the kinetic energy term is much smaller the interaction term in the Hamiltonian. This kinetic energy term can be important in the consideration of the hydrodynamics mode which is not under consideration in this paper because we have picked a fixed phase for the pair function. Thus it is reasonable to treat this kinetic energy term as a perturbation. In this approximation, the effect of a trap is to change the local condensate density at different distances from the trap center. At different positions inside the trap, we can then apply the results discussed above using the local density determined in the Thomas-Fermi approximation and $D$ is a spatial function that corresponds to the local condensate density. 

\begin{figure}
\includegraphics*[width=6cm,height=6cm]{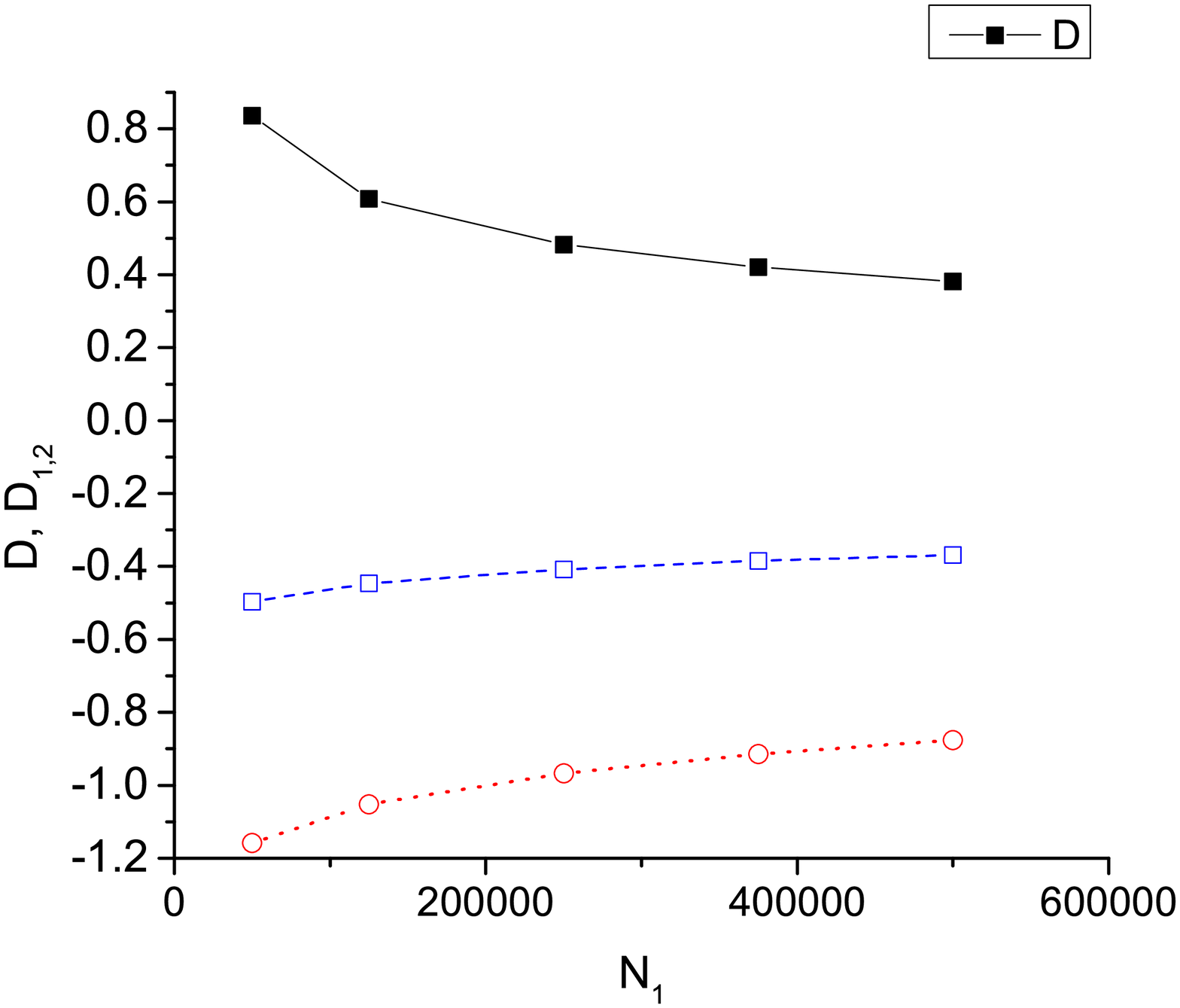}
\caption{The order parameter $D$ (black) and the parameters $D_K$ (blue $\square$) and $D_{Rb}$ (red $\circ$) in units of the trap volume density as a function of $N_K$ with $N_{Rb}=2N_K$,  $2a_{Rb-K}/a_B=-20.$} \label{ssfdn}
\end{figure}

\section{Discussion}
As we see above, the pairing state is stabilized from collapse by the intra-atomic repulsion. A critical state is reached for the case when the number of atoms the two components are the sane and the attraction is equal to the repulsion.  An example of such a state is the two component charged bosons. This particular case has been discussed by Dyson\cite{dyson} who found that the ground state energy $E<-AN^{7/5}$ for a positive constant A.


We next discuss possible ways to detect this state. From the CCD image of the expanded cloud, we can determine the density $\rho_i(\theta)$ of the particles of component i moving in a direction specified by some angular coordinates which we denote as $\theta$. We can determine a measure of the order parameters $D$ from the density correlations of particles moving in opposite directions: 
$$C = <\rho_1(\theta)\rho_2(\theta+\pi)>-<\rho_1><\rho_2>.$$ In our approximation $C\propto D^2.$

To summarise, we have described a state with both pairing order and Bose condensates for a Boson mixture with inter-component attraction. Its physical properties are described. 
In this paper we perform mean field calculations which we hope are simple enough that the physical picture is clear.  For the mean field approximation to work, the "pair size" $\xi$ need to be larger than the interparticle spacing $a_0$. We found that for our choice of parameters, $\xi/a_0\approx 3$ away from the instability point. As the instability point is approached, $\Delta$ decreases and $\xi/a_0$ increases. The mean field approximation gets better.

\section{Appendix}
If we add and subtract the two rows of the eigenvalue equation, we get
$$
H_+(c+d)=\lambda(c-d),
\\ H_-(c-d)=\lambda(c+d)
$$
where $H_{\pm}=H_1\pm H_2$ are symmetric matricies.
Taking the product, we get
$$
H_-H_+(c+ d)=\lambda^2(c+ d)
$$

This is a 2x2 equation that can be easily solved.
In the long wavelength limit  we obtain excitations with eigenvalues $\lambda^2=\Delta^2$ where the energies 
$\Delta^2$ are given by
\begin{equation}
\Delta_{\pm}^2=(-\alpha\pm \beta^{1/2})/2.
\label{eqgap}
\end{equation}

$\alpha=s_1^2 + s_2^2 + 2 u^2-2w^2  + 2 s_1 v_1 + 2 s_2 v_2,$ 
$\beta=r_1w^2+r_2w+r_3,$
$r_1=4(s_1-s_2)^2-8v_1(s_1-s_2)+8v_2(s_1-s_2)+16v_1v_2,$ 
$r_2=-16 u( s_2  v_1 + s_1  v_2 +2  v_1 v_2),$
$r_3=(s_1^2-s_2^2)^2+ 
4 (s_1 v_1 - s_2 v_2) (s_1^2 - s_2^2 + s_1 v_1 - s_2 v_2)
+4u^2 [ (s_1^2+s_2)^2+ 2 (s_1 +  s_2) (v_1 +  v_2) + 4v_1 v_2 ].$  



\begin{thebibliography}{12}
\expandafter\ifx\csname natexlab\endcsname\relax\def\natexlab#1{#1}\fi
\expandafter\ifx\csname bibnamefont\endcsname\relax
  \def\bibnamefont#1{#1}\fi
\expandafter\ifx\csname bibfnamefont\endcsname\relax
  \def\bibfnamefont#1{#1}\fi
\expandafter\ifx\csname citenamefont\endcsname\relax
  \def\citenamefont#1{#1}\fi
\expandafter\ifx\csname url\endcsname\relax
  \def\url#1{\texttt{#1}}\fi
\expandafter\ifx\csname urlprefix\endcsname\relax\def\urlprefix{URL }\fi
\providecommand{\bibinfo}[2]{#2}
\providecommand{\eprint}[2][]{\url{#2}}
\bibitem{No} See, for example, Nozieres and Saint James, J. Physique
43, 1133, 1982 and reference therein.
\bibitem{Hulet}
C. C. Bradley, C. A. Sackett, and R. G. Hulet, Phys. Rev. Lett.78, 985 (1997).
\bibitem{Mahan} G.D. Mahan, "Many Body Physics", see section 10.1 on the "Pairing Theory".
\bibitem{phase}
Franco Dalfovo and Stefano Giorgini
Lev P. Pitaevskii
Sandro Stringari
Reviews of Modern Physics, 71, 463 (1999)
seciton IVB.

\bibitem{FW}
See, for exmaple, Eq. (19.12) in "Quantum theory of Many particle systems", A. Fetter and J. D. Walecka, McGraw Hill, N.Y., (1971).

\bibitem{Oost01}  D. van Oosten, P. van der Straten and H.T.C. Stoof, Phys.
Rev. A {\bf 63}, 053601 (2001).
\bibitem{dyson} Freeman J. Dyson, Jour. Math. Phys. 8, 1538 (1967).

\end{thebibliography}
\end{document}